\documentstyle[preprint,aps]{revtex}
\input epsf.tex
\def\DESepsf(#1 width #2){\epsfxsize=#2 \epsfbox{#1}}
\begin{document}
\preprint{\vbox{\hbox{OITS-593}}}
\draft
\title{A STUDY OF SOME METHODS FOR MEASURING\\ CKM CP VIOLATING PHASES}
\author{N.G. Deshpande, Xiao-Gang He and Sechul Oh}
\address{Institute of Theoretical Science\\
University of Oregon\\
Eugene, OR 97403-5203, USA}
\date{November, 1995}
\maketitle
\begin{abstract}
We study the influence of penguin (especially, electroweak penguin) effects on
some
methods of measuring the angles $\alpha$, $\beta$, and $\gamma$ in the CKM
unitarity triangle.
We use next-to-leading order effective Hamiltonian, and present numerical
estimates based on the
factorization approximation.
We find that some techniques suggested in the literature, especially for
$\alpha$ determination,
are not workable in light of the electroweak penguin effects.  Nevertheless,
there are methods
that would work for each angle determination.
For angle $\beta$ we consider $B \rightarrow D^{+} D^{-}$ mode and estimate the
penguin
contamination.  For angle $\gamma$ we consider a method based on SU(3) symmetry
and carefully
consider SU(3) breaking effects.  We point out regions in the parameter space
where this
method could be used reliably.
\end{abstract}
\pacs{PACS numbers:11.30.Er, 12.15.Hh, 13.25.Hw}

\newpage
\section{Introduction}

The measurement of the $\epsilon$-parameter in the $K^0-\bar K^0$ meson system
is the only  direct
evidence for CP violation in the laboratory\cite{1}.  Many models have been
proposed to
explain this phenomena\cite{2,3}. The Standard Model (SM) of three generations
with the source
for CP violation arising from the phases in the Cabibbo-Kobayashi-Maskawa (CKM)
matrix is consistent
with the experiment\cite{2}. It is necessary to perform more experiments to
find out the source or
sources of CP violation and to test the CKM model. A unique feature of the CKM
model for CP
violation is that the CKM matrix is a $3\times3$ unitary matrix. Due to the
unitarity property,
when summed over the row or column of matrix elements $V_{ij}$ times complex
conjugate matrix
elements $V_{ik}^*$, the following equation holds,
\begin{eqnarray}
\sum_i V_{ij}V^*_{ik} = \delta_{jk}\;.
\end{eqnarray}
This equation defines a triangle when $j \neq k$.
For example, for $j = d$ and $k = b$ a triangle shown in Fig. 1 with three
angles
$\alpha \equiv \mbox{Arg}(-V_{td}V^*_{tb}/V_{ub}^*V_{ud})$,
$\beta \equiv \mbox{Arg}(-V_{cd}V^*_{cb}/V_{tb}^*V_{td})$, and
$\gamma \equiv  \mbox{Arg}(-V_{ud}V^*_{ub}/V_{cb}^*V_{cd})$ is defined.
These angles are related to phases of the CKM matrix elements. In Wolfenstein
parametrization the
three angles of the triangle are given by $\alpha =$ Arg $(- V_{td}/
V^{*}_{ub})$,
$\beta =$ Arg $V^{*}_{td}$, $\gamma =$ Arg $V^{*}_{ub}$\cite{4}.
The sum of these three angles must be equal to $180^0$.
This is a unique feature of the CKM model for CP violation. This property
provides an important way
to check the validity of the CKM model if enough independent measurements of
the sides and angles of
the triangle can be performed experimentally.  $B$ meson decays provide a
fertile ground to carry
out such a test\cite{5}.

 Many methods have been suggested for measuring $\alpha$, $\beta$ and $\gamma$
using $B$
decays\cite{5,6,7,8,9,10}. One class of methods involve
the measurements of CP asymmetries in time evolution of $B^0$ decays into CP
eigenstates. Such
methods make it possible to measure the three angles of the unitarity triangle
independently  and
without hadronic uncertainties if amplitudes depending on a single CKM phase
dominate the decay
process.  For instance\cite{5}, $\sin{2 \alpha}$, $\sin{2 \beta}$, and $\sin{2
\gamma}$ can be
measured in decays $B^0 \rightarrow \pi^{+} \pi^{-}$, $B^0 \rightarrow \Psi
K^0_{S}$, and $B^0_{s}
\rightarrow \rho K^0_{S}$, respectively.  However, in practice due to
simultaneous contributions to
these decays from the tree and loop (penguin) effects, the measurements are
much more difficult.
In many cases the CKM phases can not be extracted just from these asymmetries.
Additional
information is needed.  Some methods have been developed to extract CKM phases
using relations
based on isospin or flavor SU(3) symmetries to isolate the CKM phase of one
type of amplitude such
that the CKM phase determined is not contaminated by the presence of other
amplitudes\cite{7,8,9,10,11,12,13,14,15,16}. Some of these relations when used
for charged $B$
meson decay modes, also provide new methods to measure some of the CKM phase
angles.

 There are two types of penguin contributions, the strong and the electroweak
penguins. Naively, one
would expect that the electroweak penguin effects are suppressed by a factor of
$\alpha_{em}/\alpha_s$ compared with the strong penguin effects, and therefore
can be negelected.
Many previous methods for measuring the CKM phases have explicitly made such
assumption. In a
recent paper by two of us\cite{12}, we pointed out that this assumption turns
out to be
wrong for large top quark mass. Some of the methods proposed in the literature
become invalid when
the electroweak penguin effects are included. In this paper we study the
effects of the electroweak
penguin on several methods for measuring the CKM phases in the literature. We
will concentrate on
methods based on $B$ decays. There are methods based on $B_s$ decays, which
are much more difficult to perform experimentally, and we will not discuss them
in this
paper.

The paper is organized as following: In section II we present the full
effective Hamiltonian
responsible for $B$ decays, and some isospin and SU(3) analysis of the decay
amplitudes;
In section III, IV and V  we study the influence of penguin effects on some
methods for measuring
the CKM phases $\alpha$, $\beta$ and $\gamma$, respectively; And in section VI
we present our
conclusions.

\section{The effective Hamiltonian for $B$ meson decays}

 The effective Hamiltonian up to one loop level in electroweak interaction for
hadronic $B$ decays
can be written as
\begin{eqnarray}
 H_{\Delta B =1} = {4 G_{F} \over \sqrt{2}} [V_{ub}V^{*}_{uq} (c_1 O^{u}_1 +c_2
O^{u}_2)
   + V_{cb}V^{*}_{cq} (c_1 O^{c}_1 +c_2 O^{c}_2)
   - V_{tb}V^{*}_{tq} \sum_{i=3}^{12} c_{i} O_{i}] + H.C. ,
\end{eqnarray}
where $O_{i}$'s are defined as
\begin{eqnarray}
 &O&^{f}_{1} = \bar q_{\alpha} \gamma_{\mu} L f_{\beta} \bar f_{\beta}
\gamma^{\mu} L b_{\alpha} ,
  \ \  O^{f}_{2} = \bar q \gamma_{\mu} L f \bar f \gamma^{\mu} L b ,
\nonumber \\
 &O&_{3(5)} = \bar q \gamma_{\mu} L b \Sigma \bar q^{\prime} \gamma^{\mu} L(R)
q^{\prime} ,
  \ \ O_{4(6)} = \bar q_{\alpha} \gamma_{\mu} L b_{\beta} \Sigma \bar
q^{\prime}_{\beta}
       \gamma^{\mu} L(R) q^{\prime}_{\alpha}  ,
\nonumber \\
 &O&_{7(9)} = {3 \over 2} \bar q \gamma_{\mu} L b \Sigma e_{q^{\prime}} \bar
q^{\prime}
       \gamma^{\mu} R(L) q^{\prime} ,
  \ \ O_{8(10)} ={3 \over 2} \bar q_{\alpha} \gamma_{\mu} L b_{\beta} \Sigma
e_{q^{\prime}}
       \bar q^{\prime}_{\beta} \gamma^{\mu} R(L) q^{\prime}_{\alpha} ,
\nonumber \\
&O&_{11} ={g_{s}\over{32\pi^2}}m_{b}\bar q \sigma_{\mu \nu} RT_{a}b
G_{a}^{\mu \nu} \;,\;\;
Q_{12} = {e\over{32\pi^2}} m_{b}\bar q \sigma_{\mu \nu}R b
F^{\mu \nu} \;,
\end{eqnarray}
where $L(R) = (1 \mp \gamma_5)/2$, $f$ can be $u$ or $c$ quark, $q$ can be $d$
or $s$ quark,
and $q^{\prime}$ is summed over $u$, $d$, $s$, and $c$ quarks.  $\alpha$ and
$\beta$ are
the color indices.
$T^{a}$ is the SU(3) generator with the normalization $Tr(T^{a} T^{b}) =
\delta^{ab}/2$.
$G^{\mu \nu}_{a}$ and $F_{\mu \nu}$ are the gluon and photon field strength,
respectively. $c_i$
are the Wilson Coefficients (WC).  $O_1$, $O_2$ are the tree level and QCD
corrected operators.
$O_{3-6}$ are the gluon induced strong penguin operators.  $O_{7-10}$ are the
electroweak penguin
operators due to $\gamma$ and $Z$ exchange, and ``box'' diagrams at loop level.
The operators
$O_{11,12}$ are the dipole penguin operators.

The WC's $c_{i}$ at a particular scale $\mu$ are obtained by first calculating
the WC's at $m_{W}$
scale and then using the renormalization group equation to evolve them to
$\mu$.
We have carryed out this analysis using the next-to-leading order QCD corrected
WC's following
Ref.\cite{17}.  Using $\alpha_{s}(m_{Z}) =0.118$, $\alpha_{em}(m_{Z}) =1/128$,
$m_{t}=176$ GeV and
$\mu \approx m_{b}=5$ GeV, we obtain from top-quark contribution\cite{18}
\begin{eqnarray}
 c_1 &=& -0.3125,  \ \  c_2 = 1.1502,  \ \  c_3 = 0.0174,  \ \  c_4 = -0.0373,
\nonumber \\
 c_5 &=& 0.0104,  \ \  c_6 = -0.0459,  \ \  c_7 = -1.050 \times 10^{-5},
\nonumber \\
 c_8 &=& 3.839 \times 10^{-4},  \ \  c_9 = -0.0101,  \ \  c_{10} = 1.959 \times
10^{-3}.
\end{eqnarray}
It is interesting to note that the coefficient $c_9$ arising from electroweak
penguin comtribution
is not much smaller than coefficients of the strong penguin.  This enhancement
is caused by
a term in the electroweak penguin contributions in which the WC is proportional
to the square of
the top quark mass due to Z exchange.  In some application we will need
absorptive parts of $c$
and $u$ loop contributions.  These are given in Ref.\cite{18}.

The coefficients for the dipole penguin operators at the two loop level have
the following
values\cite{19}:
\begin{eqnarray}
c_{11} &=& -0.299\;, \;\;c_{12} = - 0.634\;.
\end{eqnarray}

To study exclusive $B$ decays, we also need to transform the quark operators
into hadrons. This is
a difficult task. At present there is no reliable way to carry out this
calculation. Nevertheless,
many models and suggestions have been made to provide some handle on the
related hadronic matrix
elements. Symmetry considerations provide very powerful constraints on the
matrix elements and
relate different decay amplitudes.  Isospin and flavor SU(3) symmetry are two
very useful
symmetries used in the analysis in this paper.

We can always parametrize the decay amplitude of $B$ that arises from quark
subprocess
$b \rightarrow u \bar u q$ as
\begin{eqnarray}
\bar A = <final\;state|H_{eff}^q|B> = V_{ub}V^*_{uq} T(q) +
V_{tb}V^*_{tq}P(q)\;,
\end{eqnarray}
where $T(q)$ contains the $tree$ as well as $penguin$ contributions, while
$P(q)$ contains purely
$penguin$ contributions.

When q is fixed, isospin symmetry relates some of the decay amplitudes
generated by the effective
Hamiltonian. In the case for $q=d$, isospin symmetry relates decay amplitudes
for different
$B \rightarrow \pi\pi$ or $B\rightarrow \rho \pi$ decays.
It also gives information on which operators in the effective Homiltonian
contribute to certain
isospin decay amplitudes.  This is very important for our discussions in the
rest of the paper.

For $q=d$, the tree operators $O_{1,2}$ and the electroweak penguin operators
$O_{7-10}$ contain
$\Delta I = 1/2$ and $\Delta I = 3/2$ interactions whereas the strong operators
$O_{3-6}$ and the
dipole penguin operators $O_{11,12}$ contain only $\Delta I = 1/2$ interaction.
In the case of $B\rightarrow \pi\pi$, Bose symmetry requires $\pi\pi$ to be in
$I = 0$ or $I=2$
state. Since $B$ meson is a $I=1/2$ state, we immediately know that the strong
and dipole penguin
operators will not contribute to $I =2$ decay amplitude, for example
$B^-\rightarrow \pi^- \pi^0$.
In the case for $B\rightarrow \rho \pi$, the final states can have $I =0$, 1 or
2 states.
We also easily see that the strong and dipole penguin operators will not
contribute to the
$I = 2$ decay amplitude.

For $q=s$. The tree operators $O_{1,2}$ and the electroweak penguin operators
$O_{7-10}$ contain
$\Delta I = 0$ and 1, and the strong and dipole penguin operators $O_{3-6}$ and
$O_{11,12}$
contain $\Delta I = 0$ interaction only. In $B\rightarrow \pi K$ decays, the
combinations,
$\bar A(B^-\rightarrow \pi^0 K^-) + \bar A(\bar B^0\rightarrow \pi^0 \bar K^0)$
and $ \bar
A(\bar B^0\rightarrow \pi^+ K^-) - \bar A( B^-\rightarrow \pi^- \bar K^0)$
contain only $I = 3/2$.
We immediately know that the strong and dipole penguin will not contribute to
these amplitudes.

Isospin symmetry will not relate the amplitudes between the amplitudes with
$q=d$ and amplitudes
with $q=s$. However, if one enlarges the symmetry group to the flavor SU(3),
these can be related.
The isospin relations will still be maintaned because isospin is a subgroup of
flavor SU(3)
symmetry. We shall now use SU(3) symmetry to obtain some relations which will
be
used in our later discussions.

SU(3) relations for $B$ decays have been studied by several
authors\cite{13,14,20,21}.
The operators $O_{1,2}$, $O_{3-6, 11,12}$, and $O_{7-10}$ transform under SU(3)
symmetry as $\bar 3_a + \bar 3_b +6 + \overline {15}$,
$\bar 3$, and $\bar 3_a + \bar 3_b +6 + \overline {15}$, respectively. In
general, we can
write the SU(3) invariant amplitude for $B$ decay to two octet pseudoscalar
mesons.
For the $T$ amplitude, for example, we have
\begin{eqnarray}
T&=& A_{(\bar 3)}^TB_i H(\bar 3)^i (M_l^k M_k^l) + C^T_{(\bar 3)}
B_i M^i_kM^k_jH(\bar 3)^j \nonumber\\
&+& A^T_{(6)}B_i H(6)^{ij}_k M^l_jM^k_l + C^T_{(6)}B_iM^i_jH(6
)^{jk}_lM^l_k\nonumber\\
&+&A^T_{(\overline {15})}B_i H(\overline {15})^{ij}_k M^l_jM^k_l +
C^T_{(\overline
{15})}B_iM^i_j
H(\overline {15} )^{jk}_lM^l_k\;,
\end{eqnarray}
where $B_i = (B^-, \bar B^0, \bar B^0_s)$ is a SU(3) triplet, $M_{i}^j$ is the
SU(3) pseudoscalar
octet, and the matrices H represent the transformation properties of the
operators $O_{1-12}$.
$H(6)$ is a traceless tensor that is antisymmetric on its upper indices, and
$H(\overline {15} )$ is also a traceless tensor but is symmetric on its upper
indices.
We can easily see that the strong and dipole penguin operators only contribute
to
$A_{3}$ and $C_{3}$.

For $q=d$, the non-zero entries of the H matrices are given by
\begin{eqnarray}
H(\bar 3)^2 &=& 1\;,\;\;
H(6)^{12}_1 = H(6)^{23}_3 = 1\;,\;\;H(6)^{21}_1 = H(6)^{32}_3 =
-1\;,\nonumber\\
H(\overline {15} )^{12}_1 &=& H(\overline {15} )^{21}_1 = 3\;,\; H(\overline
{15} )^{22}_2 =
-2\;,\;
H(\overline {15} )^{32}_3 = H(\overline {15} )^{23}_3 = -1\;.
\end{eqnarray}
For $q = s$, the non-zero entries are
\begin{eqnarray}
H(\bar 3)^3 &=& 1\;,\;\;
H(6)^{13}_1 = H(6)^{32}_2 = 1\;,\;\;H(6)^{31}_1 = H(6)^{23}_2 =
-1\;,\nonumber\\
H(\overline {15} )^{13}_1 &=& H(\overline {15} ) ^{31}_1 = 3\;,\; H(\overline
{15} )^{33}_3 =
-2\;,\;
H(\overline {15} )^{32}_2 = H(\overline {15} )^{23}_2 = -1\;.
\end{eqnarray}
In terms of the SU(3) invariant amplitudes, the decay amplitudes $T(\pi\pi)$,
$T(\pi K)$ for
$\bar B^0 \rightarrow \pi \pi$, $\bar B^0 \rightarrow \pi K$ are given by
\begin{eqnarray}
T(\pi^+\pi^-) &=& 2A^T_{(\bar 3)} +C^T_{(\bar 3)}
-A^T_{(6)} + C^T_{(6)} + A^T_{(\overline {15} )} + 3 C^T_{(\overline {15}
)}\;,\nonumber\\
T(\pi^0\pi^0) &=& {1\over \sqrt{2}} (2A^T_{(\bar 3)} +C^T_{(\bar 3)}
-A^T_{(6)} + C^T_{(6)} + A^T_{(\overline {15} )} -5 C^T_{(\overline {15}
)})\;,\nonumber\\
T(\pi^-\pi^0) &=& {8\over \sqrt{2}}C^T_{(\overline {15} )}\;,\nonumber\\
T(\pi^-\bar K^0) &=& C^T_{(\bar 3)}
+A^T_{(6)} - C^T_{(6)} + 3A^T_{(\overline {15} )} -  C^T_{(\overline {15}
)}\;,\nonumber\\
T(\pi^0K^-) &=& {1\over \sqrt{2}} (C^T_{(\bar 3)}
+A^T_{(6)} - C^T_{(6)} + 3A^T_{(\overline {15} )} +7 C^T_{(\overline {15}
)})\;,\nonumber\\
T(\pi^+ K^-) &=& C^T_{(\bar 3)}
-A^T_{(6)} + C^T_{(6)} - A^T_{(\overline {15} )} + 3 C^T_{(\overline {15}
)}\;,\nonumber\\
T(\pi^0\bar K^0) &=& -{1\over \sqrt{2}} (C^T_{(\bar 3)}
-A^T_{(6)} + C^T_{(6)} - A^T_{(\overline {15} )} -5 C^T_{(\overline {15} )})\;,
\nonumber\\
T(\eta_8K^-) &=& {1\over\sqrt{6}}(-C^T_{(\bar 3)}
-A^T_{(6)} + C^T_{(6)} - 3A^T_{(\overline {15} )} +9 C^T_{(\overline {15}
)})\;,\label{tt}
\end{eqnarray}
We also have similar relations for the amplitude $P(q)$. The corresponding
SU(3) invariant amplitudes will be denoted
by $A^P_i$ and $C^P_i$.
It is easy to obtain the following relations from above:
\begin{eqnarray}
 \sqrt{2}\bar A(\bar B^0\rightarrow \pi^0\pi^0) + \sqrt{2}\bar A(B^-\rightarrow
\pi^-\pi^0) =
 \bar A(\bar B^0\rightarrow \pi^+\pi^-)\;,\nonumber\\
 \bar A(\bar B^0\rightarrow \pi^+ K^-) + \bar A(B^-\rightarrow \pi^-\bar K^0)
 + \sqrt{2} \bar A(\bar B^0\rightarrow \pi^0\bar K^0) = \sqrt{2} \bar
A(B^-\rightarrow \pi^0 K^-)\;,
  \nonumber\\
 \sqrt{2} \bar A(B^-\rightarrow \pi^0 K^-) - 2\bar A(B^-\rightarrow \pi^- \bar
K^0) = \sqrt{6} \bar
 A(B^-\rightarrow \eta_8 K^-)\;.
\label{s3}
\end{eqnarray}

One expects the hadronic matrix elements arising from quark operators to be the
same order of
magnitudes, the relative strength of the amplitudes T and P are predominantly
determined by their
corresponding WC's in the effective Hamiltonian. However, in order to
numerically compare
contributions from different operators, we have to  rely on model calculations.
In our later
analysis, when such numerical calculations are required, we will use
factorization approximation.
These numerical numbers may not be accurate, but they will serve well in
providing an idea of the
validity of certain assumptions made.

\section{Measurement of the phase angle $\alpha$}

In this section we study the electroweak penguin effects on several methods
proposed for measuring
the CKM phase angle $\alpha$.

\noindent
{\bf (1) From time dependent asymmetries in $ B\rightarrow \pi\pi$ decays}

Let us first consider the standard method to measure the CKM phase $\alpha$ in
$B\rightarrow \pi\pi$ \cite{5,7}. We present it here to set up our notations
and also to clarify
some issues.
The time-dependent rate for initially pure $B^0$ or $\bar B^0$ states to decay
into a final
CP eigenstate, for example $\pi^{+} \pi^{-}$ at time $t$ is\cite{5}
\begin{eqnarray}
 \Gamma(B^0(t) \rightarrow \pi^{+} \pi^{-}) &=& |A|^2 e^{-\Gamma t}
[{1+|\lambda |^2 \over 2}
    + {1-|\lambda |^2 \over 2} \cos{(\Delta M t)} - Im \lambda \sin{(\Delta M
t)}] , \nonumber \\
 \Gamma(\bar B^0(t) \rightarrow \pi^{+} \pi^{-}) &=& |A|^2 e^{-\Gamma t}
[{1+|\lambda |^2 \over 2}
    - {1-|\lambda |^2 \over 2} \cos{(\Delta M t)} + Im \lambda \sin{(\Delta M
t)}] ,
\label{IML}
\end{eqnarray}
where $\lambda$ is defined as
\begin{eqnarray}
 \lambda \equiv {q \over p} {\bar A \over A}
\end{eqnarray}
with $A \equiv A(B^0 \rightarrow \pi^{+} \pi^{-})$ and
$\bar A \equiv \bar A(\bar B^0 \rightarrow \pi^{+} \pi^{-})$.  Here $p$ and $q$
are given by the
relations
\begin{eqnarray}
 |B_{L,H}> = p|B^0> \pm q|\bar B^0>
\end{eqnarray}
for the two mass eigenstates $B_{H}$ and $B_{L}$. In the SM,
\begin{eqnarray}
{q\over p} = {V_{tb}^*V_{td}\over V_{tb}V_{td}^*}\;.
\end{eqnarray}

The parameter $Im\lambda$ can be determined by mesuring the coefficient in CP
asymmetry in time
evolution varying with time as a sine function.

Using the effective Hamiltonian given in section II, we can parametrize
the decay amplitude in general as
\begin{eqnarray}
\bar A = V_{ub}V_{ud}^*T + V_{tb}V_{td}^*P\;.
\end{eqnarray}
If the penguin amplitude can be neglected, we have
\begin{eqnarray}
 Im \lambda = Im ({V_{tb}^*V_{td}\over V_{tb}V_{td}^*} {V_{ub}V_{ud}^* \over
V_{ub}^*V_{ud}}) = \sin (2\alpha)\;.
\end{eqnarray}
The phase $\alpha$ can therefore be determined. Similarly one can obtain
$\alpha$ from
$\bar B^0 \rightarrow \pi^0\pi^0$.

When the pegnuin effects are included, $Im\lambda$ is not equal to
$\sin(2\alpha)$ any more.
One finds\cite{12}
\begin{eqnarray}
 \Delta \sin(2\alpha) \equiv Im\lambda -\sin (2\alpha) =-R {2\cos\delta
\sin\alpha
 +\sin(2\alpha)(R-2\cos(\delta+\alpha))\over 1+ R^2-2R\cos(\delta+\alpha)}\;,
\end{eqnarray}
where $R = |P/T|$, and $\delta$ is the relative strong rescattering phase
between the $T$ and
$P$ amplitudes. It was estimated in Ref.\cite{12} using factorization
approximation that $R = 7\%$.
Using this number it was found that the error on the determination of the phase
$\alpha$ can be
as large as $12^0$.
The error is even larger if $\bar B^0 \rightarrow \pi^0\pi^0$ is used, where
$R$ is estimated to be
$23\%$.

\noindent
{\bf (2) Gronau-London method}

When penguin effects are included, the parameter $Im\lambda$ can be
parametrized as\cite{7}
\begin{eqnarray}
Im \lambda = {|\bar A|\over |A|}\sin(2\alpha + \theta)\;.
\end{eqnarray}
The ratio $|\bar A|/|A|$ can be determined by measuring the coefficient of the
term
in  CP asymmetry in time evolution varying as cosine function at
asymmetric colliders, and also at symmetric colliders\cite{22}. If $\theta$ can
be determined
independently, the phase $\alpha$ can also be determined.
To determine $\theta$, Gronau and London\cite{7} proposed using isospin
relation
\begin{eqnarray}
\sqrt{2} \bar A(\bar B^0\rightarrow \pi^0\pi^0) +\sqrt{2} \bar A(
B^-\rightarrow \pi^-\pi^0) = \bar
A(\bar B^0\rightarrow \pi^+\pi^-)\;,
\end{eqnarray}
obtained in eq.(\ref{s3}), and similar relation for the CP-conjugate amplitudes
for
the corresponding anti-particle decays. If all the six amplitudes can be
measured, the angle
$\theta$ can be determined up to two fold ambiguity as shown in Fig.2.
In this method GL argued, because $\bar B^-\rightarrow \pi^-\pi^0$ has only
$I = 2$ amplitude in the final state to which only $\Delta I = 3/2$ interaction
in the Hamiltonian
contribute, that $\bar A(B^-\rightarrow \pi^-\pi^0)$ and $\bar A(B^+\rightarrow
\pi^+\pi^0)$ have no contribution from penguin operators and therefore $|\bar
A(B^-\rightarrow
\pi^-\pi^0)| = |A(B^+\rightarrow \pi^+\pi^0)|$ . The strong penguin only has
$\Delta I = 1/2$
interaction, so the strong penguin does not contribute to this decay. If the
electroweak penguin
effects are neglected, the equlity $|\bar A(B^-\rightarrow \pi^-\pi^0)| =
|A(B^+\rightarrow
\pi^+\pi^0)|$ is exact. The electroweak penguin actually contains $\Delta
I=3/2$ interaction,
and contributes to the decay amplitude. However, since the electroweak penguin
$<$ strong penguin
$<$ tree contribution for this process, the contribution is expected to be very
small.
An estimate based on factorization gives less than $3\%$\cite{12}.
The inclusion of the electroweak penguin effects can be safely neglected. This
method, in principle,
can determine the phase $\alpha$ at a few percent level.

\noindent
{\bf (3) Hamzaoui-Xing method}

A method to measure the phase $\alpha$ without using CP asymmetry in time
evolution has
also been proposed recently by Hamzaoui and Xing\cite{11}.
We show that this method actually fails.
Based on isospin consideration and factorization approximation, they
parametrized
the decay amplitudes for $B \rightarrow \pi \pi$ as follows:
\begin{eqnarray}
 A_{+0} &\equiv& A(B^{+} \rightarrow \pi^{+} \pi^{0}) = -{1+a \over \sqrt{2}} T
e^{i \gamma},
      \nonumber \\
 A_{+-} &\equiv& A(B^{0} \rightarrow \pi^{+} \pi^{-})
    = -T e^{i \gamma} -P_{+-} e^{i (\delta - \beta)} ,     \nonumber \\
 A_{00} &\equiv& A(B^{0} \rightarrow \pi^{0} \pi^{0})
    = -{a \over \sqrt{2}} T e^{i \gamma} +{1 \over \sqrt{2}} P_{00} e^{i
(\delta - \beta)} ,
\end{eqnarray}
where the $T$ and $P_{ij}$'s are the tree and penguin amplitudes, respectively.
$\delta$ denotes the strong relative phase between the penguin and tree
amplitudes.
The parameter $a$ denotes the color-mismatched suppressed contribution in the
tree amplitudes.
If electroweak penguin effects are neglected,
\begin{eqnarray}
 P_{+-} = P_{00} \equiv P .                                        \label{PP}
\end{eqnarray}
One then obtains
\begin{eqnarray}
 \cos (\alpha + \delta) &=& {1 \over 2 \chi} [ 1 + \chi^2 -(1+a)^2 R_{+-} ],
\nonumber \\
 \cos (\alpha - \delta) &=& {1 \over 2 \chi} [ 1 + \chi^2 -(1+a)^2 \bar R_{+-}
],    \label{COS}
\end{eqnarray}
where $\chi = P/T$, and $R_{+-}$ and $\bar R_{+-}$ are the measurable
quantities defined as
$R_{ij} = |A_{ij}|^2 / (|A_{+0}|^2 + |\bar A_{-0}|^2)$ ($\bar A$ denotes the
CP-conjugate
amplitude of $A$). The parameters $a$ and $\chi$ can also be expressed in terms
of experimental
measurables,
\begin{eqnarray}
 a &=& -2 {\bar R_{00} - R_{00}\over \bar R_{+-} - R_{+-}}\;,\nonumber\\
 \chi &=& \sqrt{ (1+a)(a R_{+-}+2 R_{00}) -a }\; .
\label{CHI}
\end{eqnarray}
Therefore, if the assumptions are correct, the phase $\alpha$ could be
determined.

 In order to obtain the above equations, a cruical assumption has been made
that the parameter
$\chi_{+-} = {P_{+-}/ T}$ is equal to $\chi_{00}  = {P_{00}/ T}$.
This equality is true only if there is no electroweak penguin contribution.
The validity of the proposed method can be checked when electroweak penguin
effects are included.
Using factorization approximation, we obtain
\begin{eqnarray}
T &=& -{G_F\over \sqrt{2}} |V_{ub}V_{ud}| (\xi c_1 + c_2) T_{\pi\pi}\;,
\nonumber\\
P_{+-}&=& {G_F\over \sqrt{2}} |V_{ub}V_{ud}| [\xi c_3 + c_4
+2(\xi c_5+c_6+\xi c_7 +c_8)X_1 +\xi c_9 +c_{10}] T_{\pi\pi}\;,
\nonumber\\
P_{00} &=&{G_F\over \sqrt{2}} |V_{ub}V_{ud}| [\xi c_3 + c_4
+2(\xi c_5+c_6{1\over 2}\xi c_7 -{1\over 2}c_8)X_2
+{3\over 2}(c_7+\xi c_8 - c_9 -\xi c_{10})\nonumber\\
& -& {1\over 2} (\xi c_9 + c_{10})] T_{\pi\pi}\;,
\end{eqnarray}
where $\xi = 1/N$ with $N$ being the number of color, and
\begin{eqnarray}
X_1 &=& {m_\pi^2\over (m_b-m_u)(m_d+m_u)}\;,\;\; X_2 ={m_\pi^2\over
2m_d(m_b-m_d)}\;,\nonumber\\
T_{\pi\pi} &=& if_\pi[f^+_{B\pi}(m_\pi^2)(m_B^2-m_\pi^2) +
f^-_{B\pi}(m_\pi^2)m_\pi^2]\;.
\end{eqnarray}
In our calculations we will use $f_\pi = 132$ MeV, and the form factors
$f^\pm_{B\pi}$  in Ref.\cite{23}. For b quark mass, we use $m_b = 5$ GeV, and
for the light quark
masses $m_u = 5.6$ MeV, $m_d = 8.7$ MeV.
We also treat $\xi$ as a free parameter and use the experimentally favoured
value
$\xi \approx 1/2$\cite{24}.
We find that for $B^0 \rightarrow \pi^0 \pi^0$ the ratio of the electroweak
penguin to the strong
penguin amplitude is quite large (about $33 \%$), while that ratio for
$B^0 \rightarrow \pi^{+} \pi^{-}$ is only $5.7 \%$.
In $B^0 \rightarrow \pi^0 \pi^0$ decay, the electroweak penguin contributions
have the opposite
sign  to the strong penguin contributions and reduce the total penguin effects.
We find the values for $\chi_{+-}$ and $\chi_{00}$ are very different,
$\chi_{+-}/\chi_{00} = 1.71 $.
Here we have neglected small contributions from the dipole penguin operators
which do not
affect the result significantly.
It is clear that the assumption $\chi_{+-} = \chi_{00}$ is badly violated, and
therefore the
method proposed in Ref.\cite{11} fails completely.

\noindent
{\bf (4) Using $B \rightarrow \rho \pi$ decays}

In Ref.\cite{8,9} it was pointed out that neutral $B \rightarrow \rho \pi$ can
be used to extract
$\alpha$ without ambiguities due to penguin contributions. It was shown in
Ref.\cite{8,9} that by
studying full Dalitz plot and time dependence for $B^0 \rightarrow \pi^{+}
\pi^{-} \pi^0$,
the amplitudes and phases of $B^0 \rightarrow \rho \pi$ decays,
$S_3 = A(B^0\rightarrow \rho^+\pi^-)$, $S_4 = A(B^0\rightarrow \rho^-\pi^+)$,
$S_5 = -2 A(B^0\rightarrow \rho^0\pi^0)$, and their CP-conjugate amplitudes can
be determined.
Isospin analysis then shows that the sum $S = S_3+S_4+S_5$
has only $I = 2$ amplitude. Therefore it arises from $\Delta I = 3/2$
interaction.
If electroweak penguin effects are neglected, $S$ has only tree contribution
which contains
the phase angle $\gamma$.
Combined with angle $\beta$ from the mixing parameter $q/p$ in $B^0$-$\bar
B^0$, and
considering coefficient of $\sin{(\Delta Mt)}$ (see eq.(\ref{IML})),
the phase $\alpha$ can be determined.

Since the electroweak penguin contains $\Delta I = 3/2$ interation,
when its effects are included, $S$ contains in addition to the tree amplitude
$S_{tree}$, also
the electroweak penguin amplitude $S_{ew}$ which has a different weak phase.
The determination of the phase $\alpha$ will therefore be contaminated. In the
factorization
approximation, we obtain
\begin{eqnarray}
 S_{tree} &=& { G_{F} \over \sqrt{2}} V^{*}_{ub} V_{ud} (1+ \xi) (c_1 + c_2)
                 (\tilde{C}+\tilde{T}) ,       \nonumber \\
 S_{ew} &=& { G_{F} \over \sqrt{2}} V^{*}_{tb} V_{td} [ 3 (\xi c_7 +c_8) X
\tilde{T}
                 - {3 \over 2} (c_7 +\xi c_8) (\tilde{C}-\tilde{T})\nonumber\\
                 &-& {3 \over 2} (1+\xi) (c_9+c_{10}) (\tilde{C}+\tilde{T}) ] ,
\end{eqnarray}
where
\begin{eqnarray}
 X &=& { m^2_{\pi} \over 2 m_{d} (m_{b} + m_{d}) } ,  \nonumber \\
 \tilde{C} &=& 2 m_{\rho} (\epsilon^{*}_{\rho} \cdot p_{\pi}) f_{\rho} f^{+}_{B
\pi}
               (m^2_{\rho}) ,  \nonumber \\
 \tilde{T} &=& - 2 m_{\rho} (\epsilon^{*}_{\rho} \cdot p_{\pi}) f_{\pi} A^{B
\rho}_0
               (m^2_{\pi}) .
\end{eqnarray}
Here we have neglected a small contribution due to annihilation effects.
Note that the strong and dipole penguin operators do not contribute to $S$.
For our numerical calculations we will use  $f_{\rho} =$ 221 MeV, and the form
factors
$f^{+}_{B \pi}$ and $A^{B \rho}_0$ calculated in Ref.\cite{23}.
We find
\begin{eqnarray}
 {|S_{ew}| \over |S_{tree}|} \approx 1.4 \% ( |V_{td}| / |V_{ub}| ) .
\end{eqnarray}
We see that the ratio of the electroweak penguin to the tree contribution is
very small and $S$ is
dominated by the tree contribution.
Therefore, this method of measuring the phase $\alpha$ is good to a few percent
level.

\noindent
{\bf (5) Using $B \rightarrow \pi K$ decays}

A method for measuring the phase angle $\alpha$ has also been proposed using
$|\Delta S| = 1$ $B$ decay processes by Nir and Quinn\cite{9}. Measurement of
CP asymmetry
in time evolution in $B^0 \rightarrow \pi^0K_S$ will be able to determine
the parameter
\begin{eqnarray}
Im \lambda &=& Im ({q\over p} {\bar A(\bar B^0\rightarrow \pi^0 K_S)\over
A( B^0\rightarrow \pi^0 K_S)})=
Im (e^{-2i(\beta+\gamma)} {e^{2i\gamma}\bar A(\bar B^0\rightarrow \pi^0K_S)
\over A( B^0\rightarrow \pi^0 K_S)});. \label{ps}
\end{eqnarray}
If penguin effects are neglected, $\bar A/ A = e^{-2i \gamma}$ and so
$Im\lambda = \sin(2\alpha)$.
For this decay
it is obviously wrong to neglect the penguin effects because the penguin
contribuitons are enhanced
by a factor of $|V_{tb}V_{ts}^*/V_{ub}V_{us}^*| \approx 50$\cite{26} compared
to the tree
contributions. Even though the WC's of the penguin operators are smaller than
the tree WC's,
the net penguin contributions may be larger than the tree contributions. In
Ref.\cite{9} a method
have been proposed to overcome difficulties associated with strong penguin
effects
by determining $e^{2i\gamma} \bar A(\bar B^0\rightarrow \pi^0
K_S)/A(B^0\rightarrow \pi^0
K_S)$ directly from isospin analysis. This method requires measurement of the
decay amplitudes, $A(\bar B^0\rightarrow \pi^0\bar K^0)$,
$\bar A(\bar B^0\rightarrow \pi^+\bar K^-)$,
$\bar A( B^-\rightarrow \pi^0\bar K^-)$,
$\bar A( B^-\rightarrow \pi^-\bar K^0)$, and their CP-conjugatey amplitudes.
The strong and dipole
penguin operators do not contribute to the follwoing combinations because they
are $I = 3/2$
amplitudes, which can also be easily seen from eq.(\ref{tt}),
\begin{eqnarray}
 \bar U &\equiv& {1 \over 2} \bar A (B^{-} \rightarrow \pi^0 K^{-})
           +{1 \over 2} \bar A (\bar B^0 \rightarrow \pi^0 \bar K^0)\;,
\nonumber \\
 \bar V &\equiv& -{1 \over 2} \bar A (B^{-} \rightarrow \pi^{-} \bar K^0)
           +{1 \over 2} \bar A (\bar B^0 \rightarrow \pi^{+} K^{-})
\;,\nonumber\\
 U &\equiv& {1 \over 2} A (B^{+} \rightarrow \pi^0 K^{+})
           +{1 \over 2} A (B^0 \rightarrow \pi^0 K^0)\;,        \nonumber \\
 V &\equiv& -{1 \over 2} A (B^{+} \rightarrow \pi^{+} K^0)
           +{1 \over 2} A (B^0 \rightarrow \pi^{-} K^{+})\;.
\end{eqnarray}
If the electroweak penguin effects are neglected,
\begin{eqnarray}
U = \bar Ue^{i2\gamma}\;,\;\; V = \bar V e^{i2\gamma}\;.
\label{ss}
\end{eqnarray}
When the eight decay amplitudes for $B\rightarrow \pi K$ are measured, using
the conditions in
eq.(\ref{ss}), the quadrilaterals in eq.(\ref{s3})
\begin{eqnarray}
 \tilde A(B^{-} \rightarrow \pi^0 K^{-})
     - {1 \over \sqrt{2}} \tilde A(B^{-} \rightarrow \pi^{-} \bar K^0)
   &=& \tilde A(\bar B^0 \rightarrow \pi^0 \bar K^0)
     + {1 \over \sqrt{2}} \tilde A(\bar B^0 \rightarrow \pi^{+}
K^{-})\;,\nonumber\\
 A(B^{+} \rightarrow \pi^0 K^{+}) - {1 \over \sqrt{2}} A(B^{+} \rightarrow
\pi^{+} K^0)
   &=& A(B^0 \rightarrow \pi^0 K^0) + {1 \over \sqrt{2}} A(B^0 \rightarrow
\pi^{-} K^{+})\;,
         \label{QUAD}
\end{eqnarray}
can be constructed with $U+V$ being a common diagonal.  Here $\tilde A$'s are
defined as
$\tilde A(B \rightarrow \pi K) \equiv \bar A(B \rightarrow \pi K) e^{2i
\gamma}$.
Once these quadrilaterals are constructed, the quantity
$e^{i2\gamma}\bar A(\bar B^0\rightarrow \pi^o K_S)/A( B^0\rightarrow \pi^0
K_S)$
can be easily determined. Combining this information with eq.(\ref{ps}), the
phase
$\alpha$ can be determined.

 However, since the electroweak penguin operators also contain $\Delta I = 1$
interaction,  we have to check the validity of the relations of eq.(\ref{ss}).
In factorization approximation we find the magnitudes of the amplitudes $U$,
$V$,
$\bar U$, and $\bar V$:
\begin{eqnarray}
 U = { G_{F} \over \sqrt{2}} {1 \over 2 \sqrt{2}} &\{& V^{*}_{ub} V_{us}
     [ 2(c_1 + \xi c_2) C^{\prime} + (\xi c_1 +c_2) (T^{\prime} +A^{\prime}) ]
\nonumber \\
    &-& V^{*}_{tb} V_{ts} [-3 (c_7 + \xi c_8 - c_9 - \xi c_{10}) C^{\prime}
      + 3 (\xi c_7 + c_8) ( Y T^{\prime} + Z A^{\prime})   \nonumber \\
      &+& {3 \over 2} ( \xi c_9 + c_{10}) (T^{\prime} + A^{\prime}) ] \} ,
\nonumber \\
 V = { G_{F} \over \sqrt{2}} {1 \over 2 \sqrt{2}} &\{& V^{*}_{ub} V_{us}
     [ (\xi c_1 +c_2) ( T^{\prime} - A^{\prime})     \nonumber  \\
    &-& V^{*}_{tb} V_{ts} [ 3 (\xi c_7 +c_8) ( Y T^{\prime} - Z A^{\prime})
      + {3 \over 2} ( \xi c_9 + c_{10}) (T^{\prime} - A^{\prime}) ] \} ,
\end{eqnarray}
where
\begin{eqnarray}
 Y &=& { m^2_{K} \over (m_{b}-m_{u}) (m_{s}+m_{u}) } ,  \nonumber \\
 Z &=& { m^2_{B} \over (m_{b}+m_{u}) (m_{s}-m_{u}) } ,  \nonumber \\
 C^{\prime} &=& i f_{\pi} [ f^{+}_{BK}(m^2_{\pi}) (m^2_{B}-m^2_{K})
          + f^{-}_{BK}(m^2_{\pi}) m^2_{\pi} ] ,   \nonumber \\
 T^{\prime} &=& i f_{K} [ f^{+}_{B \pi}(m^2_{K}) (m^2_{B}-m^2_{\pi})
          + f^{-}_{B \pi}(m^2_{K}) m^2_{K} ]  ,  \nonumber \\
 A^{\prime} &=& i f_{B} [ f^{+}_{K \pi}(m^2_{B}) (m^2_{K}-m^2_{\pi})
          + f^{-}_{K \pi}(m^2_{B}) m^2_{B} ] .
\label{CTA}
\end{eqnarray}
The amplitudes $\bar U$ and $\bar V$ have the same form as $U$ and $V$,
respectively,
except that the CP-conjugate amplitudes contain the complex conjugate CKM
matrix elements
$V_{ub} V^{*}_{us}$ and $V_{tb} V^{*}_{ts}$ . As expected the strong and dipole
penguin operators
do not contribute to U and V.  The only difference between the amplitudes ($U$
and $V$) and the
CP-conjugate cmplitudes ($\bar U$ and $\bar V$) is the  opposite sign of weak
phase angle $\gamma$.
We note that if electroweak penguin contributions are neglected, we would find
that the relation of
eq.(\ref{ss}) holds.  However, we now obtain the following ratios
\begin{eqnarray}
 {| U - \bar U e^{2i \gamma}| \over |U_{tree}|} = 166 \%  |\sin{\gamma}|
         \label{aa}
\end{eqnarray}
and
\begin{eqnarray}
 {|V - \bar V e^{2i \gamma}| \over |V_{tree}|} =  42 \%  |\sin{\gamma}|,
       \label{bb}
\end{eqnarray}
where the amplitudes $U_{tree}$ and $V_{tree}$ denote the tree contribution of
$U$ and $V$ ,
respectively.
Here we have used $m_{s} =$ 175 MeV, $f_{K} =$ 162 MeV, $f_{B} =$ 200
MeV, and the form factors calculated in Ref.\cite{23}. We conclude that the
assumption presented
in Ref.\cite{9} is invalid and so the suggested isospin analysis is unworkable.

\section{MEASUREMENT OF THE ANGLE $\beta$}

In this section we study penguin effects on the determination of the CKM phase
$\beta$.
Many methods have been suggested involving $B$ meson decay into charmed
particles.
We consider two of the most convenient experimentally.

\noindent
{\bf (1) Using $B \rightarrow \Psi K_{s}$}

The easiest way to measure $\beta$ is to measure the parameter $Im\lambda$ in
CP asymmetry of
time evolution in $B^0\rightarrow \psi K_S$\cite{5}. In this case,
\begin{eqnarray}
Im\lambda = Im ({q\over p}{\bar A(\bar B^0\rightarrow \psi K_S)\over A(
B^0\rightarrow \psi K_S)})\;.
\end{eqnarray}
The tree amplitude is proportional to $V_{cb}V_{cs}^*$, and the dominant
penguin contribution from
the internal top quark is proportional to $V_{tb}V_{ts}^*$, and so the decay
amplitude can be
parametrized as
\begin{eqnarray}
A(\bar B^0\rightarrow \psi K_S) = V_{cb}V_{cs}^* T_{\psi K} + V_{tb}V_{ts}^*
P_{\psi K}\;.
\end{eqnarray}
Using unitarity of the CKM matrix, we can rewrite the above as
\begin{eqnarray}
A(\bar B^0\rightarrow \psi K_S) = V_{cb}V_{cs}^* (T_{\psi K}-P_{\psi K}) -
V_{ub}V_{us}^*
P_{\psi K}\;.
\end{eqnarray}
The WC's involved indicate that $|T_{\psi K}|$ is much larger than $|P_{\psi
K}|$.  Also
$|V_{cb}V_{cs}^*|$ is about 50 times larger than $|V_{ub}V_{us}^*|$  from
experimental data.
We can, then, safely negelect the contribution from the term proportional to
$V_{ub}V_{us}^*$.
To a very good approximation even if the penguin (strong and electroweak)
effects are included,
$Im \lambda = Im((q/p)(V_{cb}V_{cs}^*/V_{cb}^*V_{cs})) = -\sin(2\beta)$.
$\beta$ can be
measured with an error less than a percent.

\noindent
{\bf (2) Using $B \rightarrow D^{+} D^{-}$ decay}

The same conclusion can not be drawn for method to determine $\beta$ by
measuring the parameter
$Im\lambda$ in the process $\bar B^0\rightarrow D^+ D^-$. In this case, the
decay amplitude can be
parametrized as
\begin{eqnarray}
A(\bar B^0\rightarrow D^+D^-) &=& V_{cb}V_{cd}^* T_{DD} + V_{tb}V_{td}^*
P_{DD}\nonumber\\
&=& V_{cb}V_{cd}^*(T_{DD}-P_{DD}) - V_{ub}V_{ud}^*P_{DD}\;.
\end{eqnarray}
In this case although the penguin amplitude $P_{DD}$ resulting from the top
loop is suppressed
compared with the tree amplitude
$T_{DD}$, the CKM elements involved are comparable for each term. The error
caused by penguin
effects are much larger. If we keep the pegnuin contribution, we find
\begin{eqnarray}
 Im \lambda = - {\sin(2 \beta) + 2R_{DD} \sin(3\beta) \cos{\delta} +
      R_{DD}^2 \sin(4\beta) \over 1 +R_{DD}^2 + 2R_{DD} \cos(2\beta
+\delta)}\;,
\end{eqnarray}
where $R_{DD} = |P_{DD}/T_{DD}|$, and $\delta$ is the relative strong
rescattering phase between
the tree and penguin amplitudes.

In factorization approximation we can calculate the ratio $R_{DD}$.  For the
decay $\bar B^0
\rightarrow D^{+} D^{-}$, we find
\begin{eqnarray}
 \bar A(\bar B^0 \rightarrow D^{+} D^{-})
  &=& { G_{F} \over \sqrt{2}} \{ V_{cb} V^{*}_{cd} (\xi c_1 +c_2)
 \nonumber \\
  & & \mbox{} - V_{tb} V^{*}_{td} [\xi c_3 +c_4
            +2 (\xi c_5 +c_6 +\xi c_7 +c_8) X^{\prime}
            + (\xi c_9 +c_{10})]\tilde T^{\prime} ,
\end{eqnarray}
where
\begin{eqnarray}
 X^{\prime} &=& { m^2_{D} \over (m_{b} -m_{c}) (m_{c} +m_{d})} ,  \nonumber \\
 \tilde T^{\prime} &=& i f_{D} [ f^{+}_{BD}(m^2_{D}) (m^2_{B} -m^2_{D})
          + f^{-}_{BD}(m^2_{D}) m^2_{D}] .
\end{eqnarray}
Using the effective coefficients $c_{i}$, the masses $m_{D} =1.869$ GeV,
$m_{c} =1.3$ GeV, the decay constant $f_{D} =162$ MeV, and the form factors
calculated in
Ref.\cite{23}, we obtain
\begin{eqnarray}
 R_{DD} = 0.09\;.
\end{eqnarray}
Here we have neglected a small contribution due to a u-quark loop in penguin
diagram (about $8 \%$
compared with a top-quark loop contribution) and annihilation effects. In
Fig.3. we plot
$\Delta \sin(2\beta) = Im\lambda + \sin(2\beta)$ as a function of $\beta$. For
the strong rescattering
angle $\delta$, we use the quark level estimate $\delta = 12.4^0$ by including
absorptive
contribution in the WC's. The error for certain values of $\beta$ can be quite
large. For example,
for $\beta =45^0$, the error is above $16 \%$. We also carried out calculations
with the dipole
penguin operator contribution. We again find their effects to be small.

\section{MEASUREMENT OF THE PHASE $\gamma$}

In this scetion we comment on some methods for measuring the CKM phase angle
$\gamma$. Several
different classes of method to measure $\gamma$ have been proposed. One class
involve charged $B$
decays into a charge kaon and a neutral charmed meson\cite{27}. This class of
methods is not
contaminated by penguin effects because only tree amplitude can mediate such
decays.
These should work well and we do not consider them here.
Another class of method is to use information from $B\rightarrow \pi\pi$, and
$B\rightarrow
\pi(\eta) K$. In this case both tree and penguin amplitudes contribute, and
care must be taken to
include pegnuin effects.

A method to measure $\gamma$ using these decays was first studied by Gronau,
London and Rosner
(GLR)\cite{10}. In this method the tree and strong penguin effects were
considered, but the
electroweak penguin effects were neglected.  GLR argued from isospin analysis
that the combined
amplitude $\bar A( B^-\rightarrow \pi^-\bar K^0) + \bar A(B^-\rightarrow \pi^0
K^-)$ is a pure
$I = 3/2$ amplitude, and therefore only the tree amplitude contributes.
Using SU(3) relation shown in eq.(\ref{tt}), this amplitude is found to be
equal to the decay
amplitude $(f_K/f_\pi)(V_{us}^*/V_{ud})^2\bar A(B^-\rightarrow \pi^-\pi^0)$,
which can be measured.
These three amplitudes form a closed triangle, and similarly for their
CP-conjugate amplitudes.
Using the fact that $|\bar A(B^-\rightarrow \pi^-\bar K^0)| = |A(B^+\rightarrow
\pi^+ K^0)|$
(because the tree contributions here are negnigiblely small), from the two
triangles
for the particle and anti-particle decay amplitudes, the relative
phase $2\gamma$ between $A(B^-\rightarrow \pi^-\pi^0)$ and $A(B^+\rightarrow
\pi^+\pi^0)$ can be
obtained. This is a very interesting proposal.
However it was soon pointed out by Deshpande and He\cite{12} that the inclusion
of electroweak
penguin effects invalidate this method because the electroweak contributions to
$I =3/2$ amplitude
are compareable to the tree contribution.

Other methods have been proposed to take into account the electroweak penguin
effects. Recently Gronau, Hernandez, London and Rosner (GHLR)\cite{13} showed
that the
difficulty with the electroweak penguin effects can be solved by constructing
the
quadrilaterals discussed in section III for $B\rightarrow \pi K$ decays. As is
already shown, the
way used to construct the quadrilaterals in Ref.\cite{9} is not workable.
Instead, GHLR used SU(3)
relation to relate one of the diagonal of the quadrilateral to $B_s^-
\rightarrow K^-\eta$.
This time the common side of the two quadrilaterals is chosen to be
$|A(B^+\rightarrow \pi^+ K^0)| = |\bar A(B^-\rightarrow \pi^- \bar K^0)|$.
This method is however very difficult to implement experimentally because  the
decay amplitude for
$B_s^- \rightarrow K^-\eta$ is dominated by electroweak penguin contribution
and has a very small
branching ratio\cite{28}.

A more practical method has recently been  proposed by Deshpande and
He\cite{14} using SU(3) relations between the decay amplitudes for $\Delta S
=1$ decays
$B^{-} \rightarrow \pi^{-} \bar K^{0}$, $\pi^{0} K^{-}$, $\eta K^{-}$, and
$\Delta S =0$
decay $B^{-} \rightarrow \pi^{-} \pi^{0}$ .  This method requires
the construction of the triangles obtained from eq.(\ref{s3})
\begin{eqnarray}
 \sqrt{2} \bar A(B^{-} \rightarrow \pi^{0} K^{-}) -2 \bar A(B^{-} \rightarrow
\pi^{-} \bar K^{0})
    &=& \sqrt{6} \bar A(B^{-} \rightarrow \eta_8 K^{-}) ,        \nonumber \\
 \sqrt{2} A(B^{+} \rightarrow \pi^{0} K^{+}) -2 A(B^{+} \rightarrow \pi^{+}
K^{0})
    &=& \sqrt{6} A(B^{+} \rightarrow \eta_8 K^{+}),
\label{SU3}
\end{eqnarray}
 where $\eta_8$ is the pure octet component.
Using the relation
\begin{eqnarray}
 \bar A(B^{-} \rightarrow \pi^{-} \bar K^{0}) = A(B^{+} \rightarrow \pi^{+}
K^{0}) ,
\end{eqnarray}
the following result was obtained in Ref.\cite{14}:
\begin{eqnarray}
 B - \bar B = -i 2 \sqrt{2} e^{i \delta^{T}} {|V_{us}| \over |V_{ud}|}
   |\bar A(B^{-} \rightarrow \pi^{-} \pi^{0})| \sin{\gamma} ,
      \label{BB}
\end{eqnarray}
where $B$ and $\bar B$ are the complex quantities defined as
\begin{eqnarray}
 B &=& \sqrt{2} \bar A(B^{-} \rightarrow \pi^0 K^{-}) - \bar A(B^{-}
\rightarrow \pi^{-} \bar K^0) ,
      \nonumber \\
 \bar B &=& \sqrt{2} A(B^{+} \rightarrow \pi^0 K^{+})
      - A(B^{+} \rightarrow \pi^{+} \bar K^0) ,
\end{eqnarray}
shown in Fig.4.  The angle $\delta^{T}$ denotes the strong final state
rescattering phase of
the tree amplitude of $B$ (or $\bar B$).  Thus $\mbox{sin}\gamma$ can be
determined from
eq.(\ref{BB}).
This method is free from the electroweak penguin contamination problem, and all
decays involved
have relatively large ($O(10^{-5})$)  branching ratios.  They are within the
reach of future
experiments\cite{29}.

The results given in Ref.\cite{14} hold in the exact SU(3) limit.  This
relation may be broken by
SU(3) breaking effects due to $\eta$ - $\eta^{\prime}$ mixing, the breaking
effects in form
factors and mass differences.  We now make quantitative estimates of the
influence of SU(3)
breaking effects in the factorization approximation.  In this approximation we
find
\begin{eqnarray}
 F_1 &\equiv& \sqrt{2} A(B^{-} \rightarrow \pi^{0} K^{-})
        -2 A(B^{-} \rightarrow \pi^{-} \bar K^{0})
\nonumber \\
  &=& {G_{F} \over \sqrt{2}}
    \{ V_{ub} V^{*}_{us} [(c_1 +\xi c_2) C^{\prime} + (\xi c_1 +c_2)
(T^{\prime} - A^{\prime})]
         \nonumber \\
  & & \mbox{} - V_{tb} V^{*}_{ts} [-(\xi c_3 +c_4) (T^{\prime} + A^{\prime})
              -2 (\xi c_5 + c_6) (Y T^{\prime} +Z A^{\prime})
\nonumber \\
  & & \mbox{} +2 (\xi c_7 +c_8)(2 Y T^{\prime} -Z A^{\prime})
              -{3 \over 2} (c_7 +\xi c_8 -c_9 -\xi c_{10}) C^{\prime}
\nonumber \\
  & & \mbox{} +(\xi c_9 +c_{10}) (2 T^{\prime} - A^{\prime}) ] \} ,
\nonumber \\
 F_2 &\equiv& \sqrt{6} A(B^{-} \rightarrow \eta_8 K^{-})
\nonumber \\
  &=& {G_{F} \over \sqrt{2}}
    \{ V_{ub} V^{*}_{us} [(c_1 +\xi c_2) C^{\prime \prime} + (\xi c_1 +c_2)
(T^{\prime \prime}
              - A^{\prime \prime})]          \nonumber \\
  & & \mbox{} - V_{tb} V^{*}_{ts} [(\xi c_3 +c_4) (T^{\prime \prime} -2
C^{\prime \prime}
        - A^{\prime \prime}) +2 (\xi c_5 + c_6) (-2 X^{\prime \prime} C^{\prime
\prime}
        +Y T^{\prime \prime} -Z A^{\prime \prime})            \nonumber \\
  & & \mbox{} +2 (\xi c_7 +c_8)(X^{\prime \prime} C^{\prime \prime} +Y
T^{\prime \prime}
        -Z A^{\prime \prime}) -{3 \over 2} (c_7 +\xi c_8 -c_9 -\xi c_{10})
C^{\prime \prime}
         \nonumber \\
  & & \mbox{} +(\xi c_9 +c_{10}) (C^{\prime \prime} +T^{\prime \prime} -
A^{\prime \prime}) ] \},
\end{eqnarray}
where $C^{\prime}$, $T^{\prime}$, $A^{\prime}$, $Y$ and $Z$ are given in
eq.(\ref{CTA}) and
\begin{eqnarray}
 X^{\prime \prime} &=& {m^2_{\eta_8} \over 2 m_{s} (m_{b} - m_{s})} ,
 \nonumber \\
 C^{\prime \prime} &=& i f_{\eta_8} [f^{+}_{BK} (m^2_{\eta_8}) (m^2_{B} -
m^2_{K})
             + f^{-}_{BK} (m^2_{\eta_8}) m^2_{\eta_8} ]  ,
\nonumber \\
 T^{\prime \prime} &=& i f_{K} [f^{+}_{B \eta_8} (m^2_{K}) (m^2_{B} -
m^2_{\eta_8})
             + f^{-}_{B \eta_8} (m^2_{K}) m^2_{K} ]   ,
\nonumber \\
 A^{\prime \prime} &=& i f_{B} [f^{+}_{K \eta_8} (m^2_{B}) (m^2_{K} -
m^2_{\eta_8})
             + f^{-}_{K \eta_8} (m^2_{B}) m^2_{B} ]  .
\end{eqnarray}
We note that in SU(3) limit the triangle relation eq.(\ref{SU3}) is verified.
In numerical
estimates, neglecting small contribution of the annihilation diagram, we obtain
\begin{eqnarray}
 F_1 &=& {G_{F} \over \sqrt{2}} i
        [ V_{ub} V^{*}_{us} (1.86 GeV^3)e^{i\delta_T} + V_{tb} V^{*}_{ts}
(-4.82 \times 10^{-2}
        GeV^3)e^{i\delta_P} ] \ \ , \nonumber \\
 F_2 &=& {G_{F} \over \sqrt{2}} i
        [ V_{ub} V^{*}_{us} (2.70 GeV^3)e^{i\delta_T} + V_{tb} V^{*}_{ts}
(-4.79 \times 10^{-2}
        GeV^3)e^{i\delta_P} ] \ \ .  \label{s12}
\end{eqnarray}
In the above we have inserted arbitrary strong rescattering phase in the
amplitudes.
We can estimate the phases using absorptive part in WC's which indicate small
phase for $\delta_P$
and zero phase for $\delta_T$. We however keep them as free parameters here for
covenience.
For our numerical values we have used the decay constants $f_{\eta_8 } =176$
MeV, $f_{\eta_0}
\approx f_{\eta_8}$ ( $\eta_0$ is the singlet component), and the from factors
obtained in
Ref.\cite{23}.
We see that the SU(3) breaking effects in the tree amplitude are about 30\%,
and  much smaller
effects in the penguin amplitudes.

To have an idea how large the SU(3) breaking effects on the determination of
$\gamma$ are, we
carried out an excercise by taking $F_{1,2}$ in eq.(\ref{s12}) to be the
experimental values
keeping $\gamma$ and $\delta = \delta_P -\delta_T$ as free parameters. For
given values of $\gamma$
and $\delta$, we construct a triangle and obtain the value for $|B-\bar B|$. We
then take the
calculated amplitude for $|\bar A(B^-\rightarrow \pi^-\pi^0)|$ as the measured
value, and use
eq.(\ref{BB}) to determine $|\sin \gamma|$.  For each given $\gamma$, using
eq.(\ref{BB}) we will
obtain a output $\gamma$. We will call it $\gamma'$.
Corresponding to the  cases $\gamma > \delta$ and $\gamma < \delta$, there are
two solutions
for $B - \bar B$ which arise from the two possible orientations of the two
triangles relative
to their common side.  Fig. 4. shows the two triangles used to find the
magnitude $|B- \bar B|$
for $\gamma > \delta$.
We expect that because of the SU(3) breaking effects the triangle relation will
have some
deviation.  This deviation will cause errors in determining $|\sin{\gamma}|$
and the angle
$\gamma$.
In Fig. 5. we show the errors $\Delta \gamma = \gamma - \gamma'$ for a fixed
strong phase
$\delta =12^{0}$ which we expect from quark level evaluation of absorptive
parts.  Since $\delta$
is expected to be small we focus on the case $\gamma > \delta$.
{}From Fig.5 we see that there are limited range where there is solution for
$\gamma'$ because a triangle can not be formed after breaking effects for all
given $\gamma$ and
$\delta$.
We see that the errors $\Delta \gamma$ increases as $\gamma$ increases.  For
instance,
$\Delta \gamma / \gamma$ is within about $20 \%$ for $23^{0} < \gamma <35^{0}$.
For larger values
of $\gamma$ the error is larger.
If the form factors are varied by taking a different fit, it is possible to
reduce errors.
We conclude that this method is very sensitive to SU(3) breaking effects.
More theoretical efforts to study SU(3) breaking effects are called for.
We hope Lattice calculation will provide us with useful information on the
evaluation of amplitudes.

\section{CONCLUSIONS}

We have studied the influence of penguin (especially, electroweak penguin)
contributions
for several methods to extract the angles $\alpha$, $\beta$, and $\gamma$ of
the CKM unitary
triangle.  Our calculations are based on the factorization approximation using
the general
effective Hamiltonian to the next-to-leading order and on some models for form
factors when we
have needed to obtain numerical values.  To see the sensitivity of the results
on the form factors
used, we repeated our calculations using form factors obtained in
Ref.\cite{30}.
The conclusions are not changed for all cases except the calculation in Sec.V.
Here the SU(3)
breaking effects become larger.

In the cases of measuring the phase $\alpha$ using isospin relations for
$B\rightarrow \pi\pi$\cite{7} and similarly for the method using time dependent
asymmetries in
$B \rightarrow \rho \pi$\cite{8}, even though the electroweak penguin
contributions contaminate
the results, the phase $\alpha$ can be determined with an accuracy better than
a few percent because
the electroweak penguin effects are found to be quite small.  However, for the
approach proposed in
Ref.\cite{11}, we have shown that the method is unworkable since in this case
the electroweak
penguin effects are comparable to the strong penguin effects.
Similarly for method proposed in Ref.\cite{9} to extract $\alpha$ from analysis
of
$B \rightarrow \pi K$, we found that the electroweak penguin contributions can
not be neglected
and the assumption of the analysis is again invalid.

For the measurement of $\beta$, the penguin effects are negligible when use
$B\rightarrow \psi K_S$.
We found  that penguin effects are not so small in extracting $\sin{(2 \beta)}$
 from
measurement of a CP-asymmetry in $B \rightarrow D^{+} D^{-}$. The deviation
from $\sin{(2 \beta)}$
due to penguin contributions will be over $10 \%$ if the value of $\beta$ is in
the range of
$12^0 < \beta < 62^0$.

Finally we have made quantitative estimates of SU(3) breaking effects for the
method of
measuring the angle $\gamma$ proposed in Ref.\cite{14}.  We found that the
results are very
sensitive to SU(3) breaking effects and permit extraction of $\gamma$ only in a
limited
range of parameter space.  More study of the SU(3) breaking effects on the
decay amplitudes
are called for. Recently, methods using nonet symmetry\cite{15} have been
suggested.
These methods are subject to nonet symmetry breaking, and detailed studies are
required to test
their feasibility.

\begin{figure}[htb]
\vspace{1 cm}

\centerline{ \DESepsf(CKMTriangle.epsf width 12 cm) }
\smallskip
\caption {The CKM unitarity triangle.}
\vspace{3 cm}

\centerline{ \DESepsf(Quadrilateral.epsf width 12 cm) }
\smallskip
\caption {The two quadrilaterals with a common diagonal $U+V$.  Lines $a$, $b$,
$c$, and $d$ denote
   the amplitudes $A(B^0 \rightarrow \pi^0 K^0)$, $A(B^0 \rightarrow \pi^{-}
K^{+})$,
   $A(B^{+} \rightarrow \pi^0 K^{+})$, and $A(B^{+} \rightarrow \pi^{+} K^0)$.
Similarly
   lines $a^{\prime}$, $b^{\prime}$, $c^{\prime}$, and $d^{\prime}$ denote the
corresponding
   amplitudes $\tilde A$'s.}
\vspace{3 cm}

\centerline{ \DESepsf(Dsin2beta.epsf width 12 cm) }
\smallskip
\caption {$\Delta \sin{(2 \beta)} = Im\lambda +\sin{(2 \beta)}$ versus $\beta$
for a fixed strong
   rescattering phase $\delta =12.4^0$.}
\vspace{3 cm}

\centerline{ \DESepsf(SU3Triangle.epsf width 12 cm) }
\smallskip
\caption {The triangles used to find $|B - \bar B|$ for $\gamma > \delta$.
Lines $a$, $b$, and $c$
  denote the amplitudes $2^{1/2} \bar A(B^- \rightarrow \pi^0 K^-)$,
  $\bar A(B^- \rightarrow \pi^- \bar K^0)$, and $6^{1/2} \bar A(B^- \rightarrow
\eta_8  K^-)$.
  The dashed lines are for the corresponding $B^+$ decay amplitudes.}
\vspace{3 cm}

\centerline{ \DESepsf(DGamma.epsf width 12 cm) }
\smallskip
\caption {$\Delta \gamma =\gamma - \gamma^{'}$ versus $\gamma$ for a fixed
strong rescattering phase
   $\delta =12^0$.}

\end{figure}

\end{document}